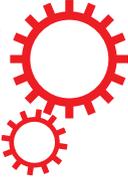

OPEN

# Renormalization group theory for percolation in time-varying networks

Jens Karschau, Marco Zimmerling & Benjamin M. Friedrich

**Motivated by multi-hop communication in unreliable wireless networks, we present a percolation theory for time-varying networks. We develop a renormalization group theory for a prototypical network on a regular grid, where individual links switch stochastically between active and inactive states. The question whether a given source node can communicate with a destination node along paths of active links is equivalent to a percolation problem. Our theory maps the temporal existence of multi-hop paths on an effective two-state Markov process. We show analytically how this Markov process converges towards a memoryless Bernoulli process as the hop distance between source and destination node increases. Our work extends classical percolation theory to the dynamic case and elucidates temporal correlations of message losses. Quantification of temporal correlations has implications for the design of wireless communication and control protocols, e.g. in cyber-physical systems such as self-organized swarms of drones or smart traffic networks.**

Renormalization group (RG) theory elegantly addresses percolation in static networks[1–3]. Percolation refers to the existence of large connected components in a random graph. Specifically, for subgraphs of a regular lattice, a giant connected component emerges above a critical lattice filling fraction, thus marking a phase transition of percolation. Percolation theory has been applied to a range of phenomena, from fluid flow in porous materials to epidemic spreading[4–8]. In this paper, we apply RG theory to time-varying communication networks.

Our work is motivated by wireless communication networks that often exhibit unreliable links. There, a key question concerns the existence of a multi-hop path of simultaneously active links, which permits sending a message from a source to a destination node via one or several intermediate relay nodes. Real-world applications of particular relevance include self-organizing swarms of flying drones[9], smart traffic networks of communicating cars[10], and networks of cooperating robots in production lines[11]. Recent flooding and multi-path routing protocols were shown to be more reliable than traditional single-path routing in field experiments[12,13]. The emergence of ever larger wireless networks that serve as critical communication infrastructures for cyber-physical applications[14] prompts the need for a theoretical understanding of message losses and their temporal correlations when using these protocols[15]. Widely used schemes to estimate the quality of a wireless link assume that message losses are uncorrelated in time[16]. But temporal correlations among losses render these estimates invalid, and hence may cause existing communication protocols and control algorithms to fail[17,18]. This question of temporal correlations of message losses falls into a recent, application-driven interest in time-varying networks[19–21].

Here, we introduce a minimal model of percolation in time-varying networks, which captures key features of multi-path wireless communication with unreliable links. Most real-world applications exhibit fairly regular network topologies, such as swarms of drones flying in a formation[9], or sensor arrays in smart production facilities. Thus, we consider the case of network nodes distributed on a regular lattice, connected by links that stochastically switch between being active or inactive with finite switching time. The case of two states per link, active and inactive, serves as illustrative example, and corresponds to, e.g. a data transmission rate of an individual link that is either above or below the threshold, which guarantees a certain quality of service. We ask for the existence of multi-hop paths consisting of simultaneously active links that connect a designated source and destination node. We assume that transmission delays are short compared to the stochastic switching time of individual links. Indeed, in low-power wireless networks, transmission times are at most a few milliseconds per link, whereas the stochastic switching times of links can be on the order of hundreds of milliseconds. Now, if one were just interested in the probability to find a multi-hop path at a single point in time, the question would reduce cfaed, TU Dresden, 01069, Dresden, Germany. Correspondence and requests for materials should be addressed to B.M.F. (email: benjamin.m.friedrich@tu-dresden.de)

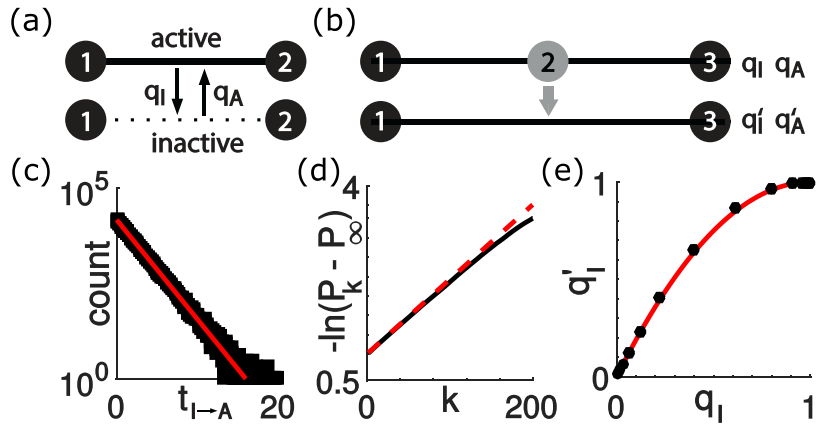

**Figure 1.** Renormalization of a two-link motif. (**a**) We consider communication links that can switch between an active and an inactive state with transition probabilities $q_I$ and $q_A$. (**b**) The switching dynamics of a two-link motif can be well approximated by a single link with effective transition probabilities $q'_I$ and $q'_A$. (**c**) Waiting times for the transition $I \to A$ are approximately exponentially distributed for the two-link motif shown in (**b**). (**d**) Comparison of simulation results (solid line) and analytical results (dashed line) for the state correlation function given by Eq. (3) with slope $-\ln \beta'$. (**e**) Comparison of simulation results (circles) and analytical results (solid line) for $q'_I$.

to a bond-percolation problem for a static network, where the probability of an individual link to be active plays the role of the lattice filling fraction. Previous mean-field descriptions have shown that this probability converges to 1 for large networks if the lattice filling fraction is above a critical value known as the percolation threshold[5,8]. However, mean-field descriptions cannot account for temporal correlations of message losses at the network level. This motivates us to extend classical percolation theory for static networks to the dynamic case by formulating an RG theory that provides effective switching rates at different coarse-graining levels of a network. Using this theory, we quantify how message losses are correlated in time. We thus assess the validity of a common practice to assume uncorrelated message losses during the design and analysis of wireless protocols and control loops[17,18].

## Model and Metrics

**A minimal model of time-varying networks.** We consider a network $(V, E)$ with nodes $V$ and undirected edges $E \subset V \times V$. Every edge represents a communication link that can be either *active* or *inactive*. Active links can relay information from one node to the next, while inactive links cannot. In each time step, a link can switch between the active and the inactive state with respective transition probabilities $q_I = q_{A \to I}$ and $q_A = q_{I \to A}$, which corresponds to a time-discrete telegraph process, see Fig. 1(a). We assume that the switching dynamics of different links is uncorrelated, which is a valid assumption for the vast majority of links in real wireless networks[18]. Our goal is to accurately represent a given network $(V, E)$ by a coarser version $(V', E')$ with fewer nodes and links, while reproducing key metrics.

**Reliability.** We define the reliability of a link as the probability $\Phi$ to find the link in the active state. In our two-state description, we have

$$\Phi = q_A/(q_I + q_A). \tag{1}$$

**Bernoulliness score.** To characterize temporal correlations between states $s_n = s(t_n)$ of a link at different times $t_n = n\Delta t$, we introduce a Bernoulliness score $\beta$

$$\beta = 1 - q_I - q_A. \tag{2}$$

The probability $P(s_{n+k} = s_n)$ to find a link in the same state after $k$ time steps is a function of the Bernoulliness $\beta$

$$P(s_{n+k} = s_n) = \Phi^2 + (1 - \Phi)^2 + 2\Phi(1 - \Phi)\beta^k. \tag{3}$$

The corresponding correlation time is $\tau = -\Delta t/\ln \beta$. For $\beta = 0$, the stochastic switching of a link becomes a memoryless Bernoulli process with $\tau = 0$, whereas for $\beta > 0$ it retains memory and must be described as a Markov process.

**Bernoulliness impacts the estimation of reliability.** Temporal correlations in the state of a link become important, e.g. in protocols that estimate the link reliability $\Phi$ by observing the state of a link at $n$ subsequent time steps. The estimate $\Phi_{\text{est}} = n_A/n$, where $n_A$ is the number of time steps for which the link is active, is itself a random variable. In the absence of temporal correlations with $\beta = 0$, the variance $\sigma^2(\Phi_{\text{est}})$ of $\Phi_{\text{est}}$ is given by

$$\sigma^2(\Phi_{\text{est}}) = \Phi(1 - \Phi)/n. \tag{4}$$

For $\beta > 0$, this result changes to

$$\sigma^2(\Phi_{est}) \approx -2\ln\beta\,\Phi(1-\Phi)/n, \tag{5}$$

corresponding to a reduced effective number $n_\beta$ of independent time steps $n_\beta = -n/(2\ln\beta)$. The derivation of Eq. (5) as well as a comparison with simulations is given in the Supplementary Material text (SM) that accompanies this paper.

## Renormlization theory of time-varying networks

### Illustrating the method: renormalization of a two-link motif.

We compute an effective Bernoulliness score for networks of arbitrary size. To illustrate the general procedure, we first consider the simple case of a chain of $n = 2$ links with a set of nodes $E = \{1, 2, 3\}$ and a set of links $V = \{(1, 2), (2, 3)\}$, see Fig. 1(b). We say this network is active if the outermost nodes 1 and 3 are connected by a path of active links. In general, the state space is $S = \{A, I\}^n$, where $n = |E|$ denotes the number of nodes. The transition probabilities $q_{s \to s'}$ for a transition from state $s \in S$ to a new state $s' \in S$ read

$$q_{s \to s'} = q_I^{n_{A \to I}}(1-q_I)^{n_{A \to A}} q_A^{n_{I \to A}}(1-q_A)^{n_{I \to I}}. \tag{6}$$

Here, the $n$'s denote the respective number of links that switch or retain their state, as indicated by the subscript, e.g., $n_{A \to I}$ denotes the number of links that switch from active (A) to inactive (I) upon a change of state $s$ to $s'$. The steady-state probability of state $s$ reads

$$P_s^* = \Phi^{n_A}(1-\Phi)^{n_I}, \tag{7}$$

which defines a probability distribution $\mathbf{P}^* = \{P_s^*\}_{s \in S}$. Here, $n_A$ and $n_I$ denote the number of active and inactive links in network state $s$, respectively.

For the example of the two-link motif, the set $S_A$ of active network states is simply $S_A = \{AA\}$. The probabilities $P_A$ and $P_I$ for the network to be active or inactive at steady state are $P_A = \sum_{s \in S_A} P_s^*$ and $P_I = \sum_{s \in S_I} P_s^*$, respectively, where $S_I = S \setminus S_A$ denotes the set of inactive network states. The probability current $J_{A \to I}$ from the active to the inactive state of the motif reads

$$J_{A \to I} = \sum_{s \in S_A} \sum_{s' \in S_I} q_{s \to s'} P_s^*, \tag{8}$$

and similarly for the probability current $J_{I \to A}$ from the inactive back to the active state. Eq. (8) is exact in the absence of temporal correlations, $\beta = 0$, and represents a valid approximation in the case of temporal correlations, where the probability distribution of network states relaxes to $\mathbf{P}^*$ on a time-scale $\tau$, with correlation time $\tau = -\Delta t/\ln\beta$ as it was introduced above. This motivates a replacement of the two-link chain by a single link with effective transition probabilities $q'_I$ and $q'_A$ with

$$q'_I = J_{A \to I}/P_A, \quad q'_A = J_{I \to A}/P_I, \tag{9}$$

see Fig. 1(b). For the two-link motif, we have $q'_I = q_I(2-q_I)$ and $q'_A = q_A^2(2-q_I)/(q_I + 2q_A)$. The renormalization map can be equivalently expressed in terms of an end-to-end reliability $\Phi'$ and an end-to-end Bernoulliness $\beta'$ of the two-link network motif

$$\Phi' = \Phi^2, \quad \beta' = \beta[\beta + (2-\beta)\Phi]/(1+\Phi). \tag{10}$$

Thus, we have approximated the Markovian dynamics with $|S| = 4$ states of the network motif by an effective two-state Markov model. Figure 1(c–e) compare these analytical results with numerical simulations, validating the applicability of this two-state approximation.

### The linear chain.

We can apply the above coarse-graining argument iteratively to a larger network, such as the linear chain of length $n$ with set of nodes $E = \{1, 2, \ldots, n+1\}$ and set of links $V = \{(1, 2), (2, 3), \ldots, (n, n+1)\}$. The only active network state that allows information transmission between the outer-most nodes 1 and $n+1$ of the chain is $AA \ldots A$. We consider the special case $n = 2^k$ for some integer $k$. By applying the renormalization map $(q_I, q_A) \to (q'_I, q'_A)$ $k$-times to a $2^k$-chain, we reduce the chain to a single link with effective transition probabilities $q_I^{[k]}$ and $q_A^{[k]}$, where the superscript $[k]$ denotes the $k$-th iterate of the renormalization map. Explicitly,

$$q_I^{[k]} = 1 - (1-q_I)^n, \quad q_A^{[k]} = q_A^n \frac{1-(1-q_I)^n}{(q_I + q_A)^n - q_A^n}. \tag{11}$$

For this simple example, we can also compute the effective transition probabilities for the $n$-link chain directly, using Eq. (9). Thereby, we recover the right-hand sides of Eq. (11) for arbitrary $n$, which validates our RG approach.

We now ask for the fixed points under the renormalization map, defined by $(q'_I, q'_A) = (q_I, q_A)$. These fixed points correspond to special values of the transition probabilities of individual links, $q_I$ and $q_A$, for which the two-link motif will have the same effective transition probabilities as its individual links. By induction, the same will hold true for the linear chain with $n = 2^k$ links. We find a trivial set of unstable fixed points with $q_I = 0$, which corresponds to the case of perfect communication links with $\Phi = 1$. Further, there is a unique stable fixed point $(q_I, q_A) = (1, 0)$, which corresponds to the case of permanently inactive links with $\Phi = 0$. The end-to-end reliability of a linear chain with $\Phi > 0$ converges to this stable fixed point as $\Phi^{[k]} = \Phi^n$ for increasing chain length $n = 2^k$. This is consistent with the intuition that longer chains become more unreliable. Concomitantly, the end-to-end Bernoulliness converges as a geometric series to zero

$$\beta^{[k]} = \beta^n + \mathcal{O}(\Phi), \quad n = 2^k. \tag{12}$$

Next, we apply a similar coarse-graining procedure to networks of node degree $d > 2$ that allow for multi-path routing.

**Triangular networks.** As an example, we consider an infinite triangular network with nodes $V = \{m\mathbf{e}_1 + n\mathbf{e}_2 \in \mathbb{R}^2 | m, n \in \mathbb{Z}\}$ on a lattice spanned by basis vectors $\mathbf{e}_1 = (1, 0)$ and $\mathbf{e}_2 = (1/2, \sqrt{3}/2)$ and corresponding set of links $E$, see Fig. 2(a). We define a sub-network $(V', E')$, whose nodes correspond to the lattice spanned by new basis vectors $\mathbf{e}'_1 = \mathbf{e}_1 + \mathbf{e}_2$ and $\mathbf{e}'_2 = 2\mathbf{e}_2 - \mathbf{e}_1$, which is shown in blue in the same figure panel. This sub-network can be equivalently obtained by considering the original lattice as a tessellation by the 4-node motif shown in Fig. 2(b) and replacing each motif by a single link. This purely geometric procedure of network coarse-graining can be repeated.

Following an analogous procedure as for the two-link motif, we compute the end-to-end reliability $\Phi'$ and end-to-end Bernoulliness $\beta'$ for the 4-node motif expressible in terms of a renormalization map

$$(\Phi', \beta') = \mathcal{R}[\Phi, \beta]. \tag{13}$$

The renormalization map expresses the metrics of the network motif in terms of the metrics of the individual links it is composed of. The full analytic expression for Eq. (13) is given in the accompanying SM. We find that there exists a critical value $\Phi_c^{RG} = 1/2$ such that the renormalized reliability decreases, $\Phi' < \Phi$, if $\Phi < \Phi_c^{RG}$, while it increases, $\Phi' > \Phi$, if $\Phi < \Phi_c^{RG}$. The critical value $\Phi_c^{RG}$ thus represents a percolation threshold: if the reliability of individual links is below this value, repeated application of the renormalization map Eq. (13) predicts a decreasing end-to-end reliability of networks of increasing size that converges to zero, whereas if the link reliability is above this value, the end-to-end reliability will converge to a perfectly reliable network. If $\Phi = 0$ or $\Phi = 1$, the analytic expression for $\beta'$ simplifies to $\beta' = \beta^2$.

Figure 2(d) summarizes the RG action in $(\Phi, \beta)$-parameter space. We can analyze this RG action using concepts from dynamical systems theory. The point $(\Phi_c^{RG}, 0)$ is a saddle node with unstable manifold given by $\beta = 0$ and stable manifold given by $\Phi = \Phi_c^{RG}$. This stable manifold serves as separatrix, separating the basins of attraction of two stable fixed points $(0, 0)$ and $(1, 0)$. Thus, repeated coarse-graining of an infinite triangular lattice will converge to a sub-lattice with perfectly transmitting links if and only if $\Phi > \Phi_c^{RG}$, consistent with classical percolation theory[22]. Moreover, our analysis shows that the switching dynamics converges to a memoryless Bernoulli process on increasingly coarse-graining scales. Generally, the renormalized correlation time $\tau^{[k]}$ after $k$ coarse-graining steps satisfies $\tau > \tau^{[k]} \geq \tau/n$, where $n = 2^k$ is the length of the shortest path between source and destination node. We have equality at the lower bound $\tau^{[k]} = \tau/n$ precisely in the limit cases of perfectly unreliable or perfectly reliable links, $\Phi = 0$ and $\Phi = 1$. For intermediate values of the link reliability $\Phi$, $0 < \Phi < 1$, the correlation time $\tau^{[k]}$ decays logarithmically as $1/\ln(n)$ in the asymptotic limit of large $n$. Interestingly, the coefficient $\alpha_1(\Phi)$ in the polynomial expression $\beta' = \sum_{i=1}^{5} \alpha_i(\Phi)\beta^i$ is maximal at $\Phi = \Phi_c^{RG}$, implying that the asymptotic decay of temporal correlations is slowest right at the percolation threshold.

Simulations corroborate this picture. Figure 2(e) shows end-to-end reliability $\Phi(M_k; \Phi, \beta)$ and Bernoulliness $\beta(M_k; \Phi, \beta)$ for iterated network motifs $M_k$ with link reliability $\Phi$ and link Bernoulliness $\beta$. Here, $M_0$ is a single link, and $M_k$ is the motif, for which a single coarse-graining step gives $M_{k-1}$, see Fig. 2(c). We have

$$\Phi(M_k; \Phi, \beta) \approx \Phi(M_{k-1}; \Phi', \beta') \approx \cdots \approx \Phi^{[k]}, \tag{14}$$

and analogously for $\beta(M_k; \Phi, \beta)$. Thus, the end-to-end reliability and Bernoulliness of iterated network motifs are approximately given by the effective reliability and Bernoulliness of individual links of iteratively coarse-grained networks. Deviations between RG theory and simulations stem from the fact that the 4-link motifs that are used to coarse-grain the network $M_{k+1}$ to the smaller network $M_k$ share common links. Consequently, switches of the effective links of $M_k$ are not completely uncorrelated to each other, as assumed in our RG calculation. The percolation threshold $\Phi_c^{RG}$ from RG theory overestimates the value $\Phi_c \approx 0.35$ for a static triangular network[23], a feature known from zeroth-order RG theory[1].

The RG approach generalizes in a straight-forward manner to the case, where the switching rates are drawn from a distribution with given mean and variance, resulting in a renormalization of the distributions, see Fig. S2 in SM.

Our theory assumes that transmission and processing delays are much shorter than the time-scale of link switching, which is a realistic assumption. If each message would spend a non-zero processing delay at each node, we find that this delay has only a minor effect on the network Bernoulliness, provided the sending interval at which messages are injected into the network at the source node stays constant, see Fig. S3 in SM.

**Other network topologies.** Our general approach can be applied in a similar manner to other networks with varying node degree $d$, including regular 1-, 2-, and 3-dimensional lattices. Table 1 summarizes the results. As coarse-graining motif, we used a 2-link motif in the case of the linear chain, a 4-link motif in the case of the triangular lattice [see Fig. 2(b)], a 33-link motif consisting of 4 unit cubes in the case of the cubic lattice (see Fig. S4 in SM), and the lattice unit cell in all other cases.

In line with classical percolation theory, the end-to-end reliability $\Phi^{[k]}$ converges to 0 with increasing number $k$ of coarse-graining steps if $\Phi < \Phi_c^{RG}$ for some critical value $\Phi_c^{RG}$ and converges to 1 otherwise. The number of multi-hop paths of given length between source and destination node increases with node degree $d$, which is reflected by lower values of the percolation threshold $\Phi_c^{RG}$. The end-to-end Bernoulliness $\beta^{[k]}$ predicted by RG theory converges to zero, yet at different speeds. For perfectly unreliable or perfectly reliable links, $\Phi = 0$ or $\Phi = 1$, $\beta^{[k]}$ decreases with a specific exponent in each coarse-graining step: we have $\lim_{\Phi \to 0} \beta^{[k]} = \beta^{ak}$, where $a$ denotes the number of links in the shortest path connecting the source and the destination node of the coarse-graining motif, and $\lim_{\Phi \to 1} \beta^{[k]} = \beta^{bk}$, where $b$ denotes the number of links starting from the source node in the coarse-graining motif (or the minimum of this number and the number of links starting from the destination node, in case these two numbers are different). Contrastingly, close to the percolation threshold $\Phi_c^{\text{RG}}$, temporal correlations decay only logarithmically, reflected by $\beta' \sim \beta$.

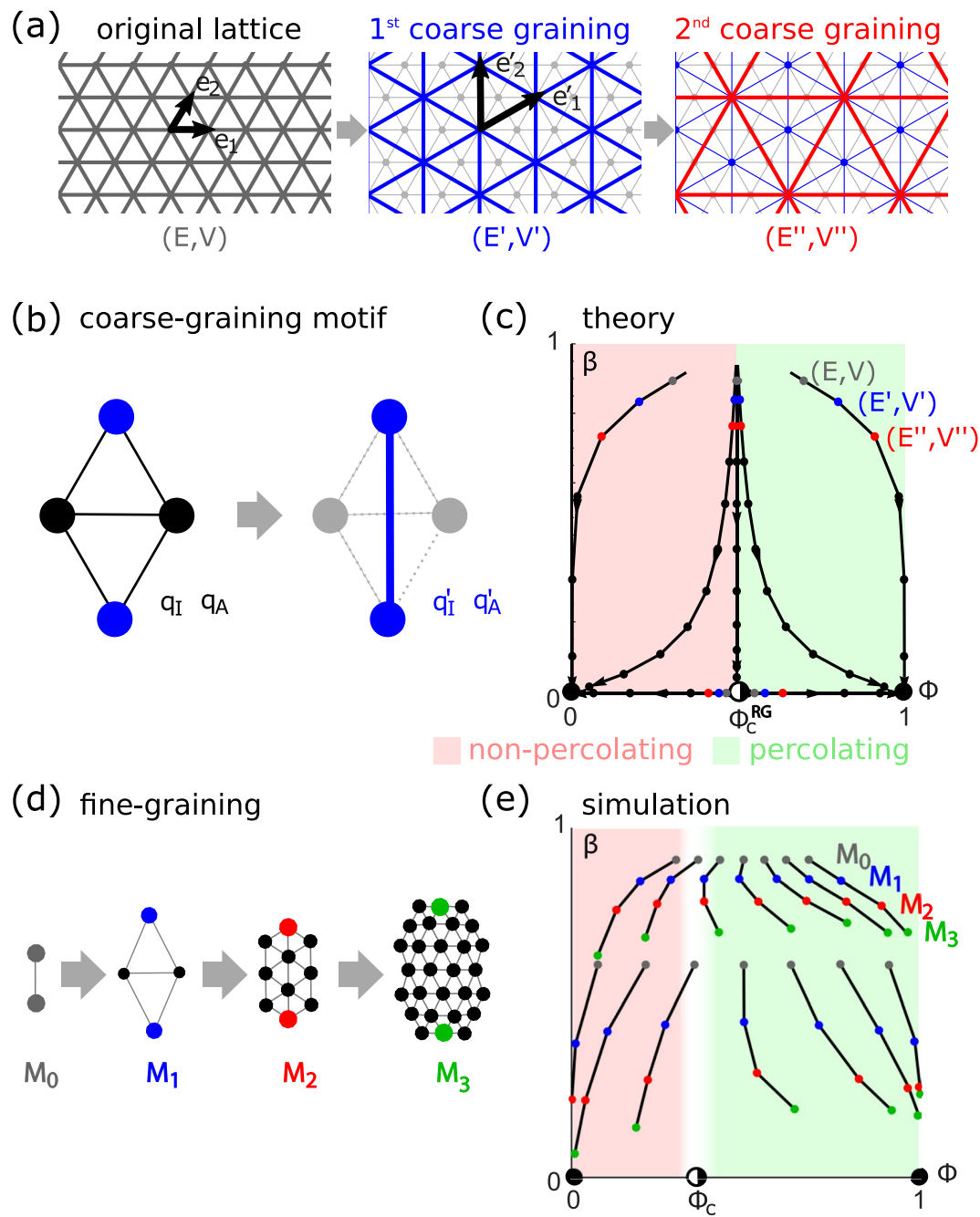

**Figure 2.** Renormalization of a time-varying network on a triangular lattice. (**a**) Subsequent stages of network coarse-graining: gray $(V, E)$, blue $(V', E')$, red $(V'', E'')$. (**b**) The 4-node motif used in the coarse-graining procedure. (**c**) Flow in $(\Phi, \beta)$-space under the action of the renormalization group showing example trajectories (black), stable fixed points (solid circles), respective basins of attraction (rosé/green), and saddle node $(\Phi_c, 0) = (1/2, 0)$ (half-filled circle). (**d**) Network iterates $M_k$ used in simulations. End-to-end reliability and Bernoulliness for iteratively fine-grained network motifs correspond to the renormalized reliability and Bernoulliness of individual effective links in coarse-grained networks, see Eq. (14). (**e**) Simulation results for end-to-end reliability $\Phi(M_k)$ and Bernoulliness $\beta(M_k)$ for motifs $M_k$.

| Lattice | $d$ | $\Phi_c$ True | $\Phi_c$ RG | $\beta'$ $\Phi=0$ | $\Phi=1$ | $\Phi=\Phi_c^{\text{RG}}$ |
|---|---|---|---|---|---|---|
| Linear chain | 2 | 1 | 1 | $\beta^2$ | — | $\beta$ |
| Honeycomb | 3 | ≈0.65 | ≈0.85 | $\beta^3$ | $\beta^2$ | $0.47\beta$ |
| Square | 4 | 1/2 | $(\sqrt{5}-1)/2$ | $\beta^2$ | $\beta^2$ | $0.58\beta$ |
| Triangular | 6 | $2\sin(\pi/18)$ | 1/2 | $\beta^2$ | $\beta^2$ | $\frac{37}{64}\beta$ |
| Cubic | 6 | ≈0.25 | ≈0.50 | $\beta^5$ | $\beta^3$ | $0.36\beta$ |

**Table 1.** Network topology determines percolation threshold and decay of temporal correlations. For different regular lattices of node degree $d$, we state bond-percolation thresholds $\Phi_c^{\text{RG}}$ from our RG theory, and known percolation thresholds $\Phi_c$[22,23,26,27]. The end-to-end Bernoulliness $\beta'$ after a single coarse-graining step for three link reliabilities $\Phi$ to leading order in $\beta$.

### Application case study: a swarm of communicating drones

As an application of our RG approach, we consider a simple model of a swarm of drones that communicate using short-range wireless radios[9]. Each drone diffusively explores a region around a fixed grid position $(x_0, y_0)$ in two-dimensional space, see Fig. 3(a). The dynamics of its time-dependent position $\mathbf{r}(t) = (x(t), y(t))$ shall follow an Ornstein-Uhlenbeck process $\gamma \dot{x}(t) = x_0 - x(t) + \xi(t)$, where $\xi(t)$ is white Gaussian noise with $\langle \xi(t)\xi(t') \rangle = a^2 \gamma \delta(t-t')$, and analogous dynamics for the $y$ coordinate. This model matches a physical scenario where swarms of drones fly in regular formation with fluctuations in position due to, e.g. wind gusts and navigation imperfections. We assume that a pair of drones can communicate with a dynamic probability $f(r)$ that is a function of the Euclidean distance $r = |\mathbf{r}_i - \mathbf{r}_j|$ between the current positions $\mathbf{r}_i$ and $\mathbf{r}_j$ of the two drones. Indeed, this is a valid assumption in free-space aerial environments[24]. We choose a Hill function, $f(r) = 1/[1 + (r/r_0)^h]$, with Hill coefficient $h$ and half-saturation constant $r_0$. We define an effective link reliability $\Phi$ as the time-averaged communication probability for a pair of drones at neighboring grid positions, $\Phi = \langle f(r) \rangle$. A designated source drone sends messages in regular intervals of $t_{\text{send}}$.

The correlation time for end-to-end communication along a single multi-hop path from simulations compares favorably to RG predictions based on values for a single link, see Fig. 3(c). For swarm motifs $M_k$ of increasing size $k$, we compute end-to-end reliability and Bernoulliness, see Fig. 3(d). We observe again key features predicted by RG flow: a percolation transition and slow reduction of temporal correlations.

Hence, for a routing protocol that uses flooding regions $M_k$ as suggested by our coarse-graining procedure, we find that transmission will be highly reliable provided two conditions are met (i) the network must be above the percolation threshold, $\Phi > \Phi_c$, and (ii) the distance between source and destination node is sufficiently large, allowing for multiple renormalization steps. We now present a modified routing protocol that relaxes the second condition by enlarging the flooding region, thereby increasing the number of potential multi-hop paths. Specifically, exactly those drones shall relay the signal whose reference positions are within a distance $M$ of the shortest path between the source and the destination node, see cartoon in Fig. 3(e).

Depending on whether the network is above or below the percolation threshold, the end-to-end reliability $\Phi_M$ exhibits a qualitatively different behavior as function of $M$, see Fig. 3(e): For effective link reliabilities above the percolation threshold $\Phi > \Phi_c$, the end-to-end reliability $\Phi_M$ for flooding regions of increasing width $M$ converges to 1, corresponding to reliable end-to-end transmission, while for link reliabilities above the percolation threshold $\Phi < \Phi_c$, the end-to-end reliability $\Phi_M$ converges to a fixed point $\Phi_\infty$ with $\Phi_\infty < 1$. Remarkably, the end-to-end Bernoulliness $\beta_M$ converges to a non-zero value $\beta_\infty$, even for $\Phi > \Phi_c$, in this case of a fixed distance between source and destination node. This highlights the role of hop distance for the decay of temporal correlations. We can ask for the minimal width $M_{95\%}$ of flooding regions that ensures a specified end-to-end reliability, here chosen as 95% based on typical application requirements[25], see Fig. 3(g). Remarkably, for effective link reliabilities $\Phi$ in a range [0.6, 1], $M_{95\%}$ takes moderate values of 2–3, which are largely independent of both link reliability and the distance between source and destination node, thus suggesting a simple multi-path routing protocol of restrained flooding for robust wireless multi-hop communication.

### Discussion

Here, we presented a percolation theory for time-varying networks, inspired by multi-hop communication in wireless networks with unreliable links. We introduced a minimal model of networks on regular lattices, where individual network links switch stochastically between active and inactive states. While active links can relay messages from one node to the next, message loss occurs at inactive links. With the help of this minimal model, we address the fundamental question of how temporal correlations of message losses for large communication networks scale with network size. To quantify these temporal correlations, we introduce a new metric of Bernoulliness. Our analytical theory allows to calculate end-to-end reliability and end-to-end Bernoulliness for networks of arbitrary size by repeated application of a renormalization map.

Full simulations of large time-varying networks are computationally expensive because long simulation times are required to accurately estimate network metrics. In contrast, the computational complexity of our approximative method to compute network metrics for large networks is independent of the size of the network, as the calculation only involves repeated application of the renormalization map, which is computationally cheap. The computational complexity for pre-computation the renormalization map scales at most as $n(2^n)^2$ if a network motif with $n$ links is used as renormalization motif. Note here that we used the worst-case complexity of Dijkstra's algorithm, which determines whether a given state of the motif is active or inactive.

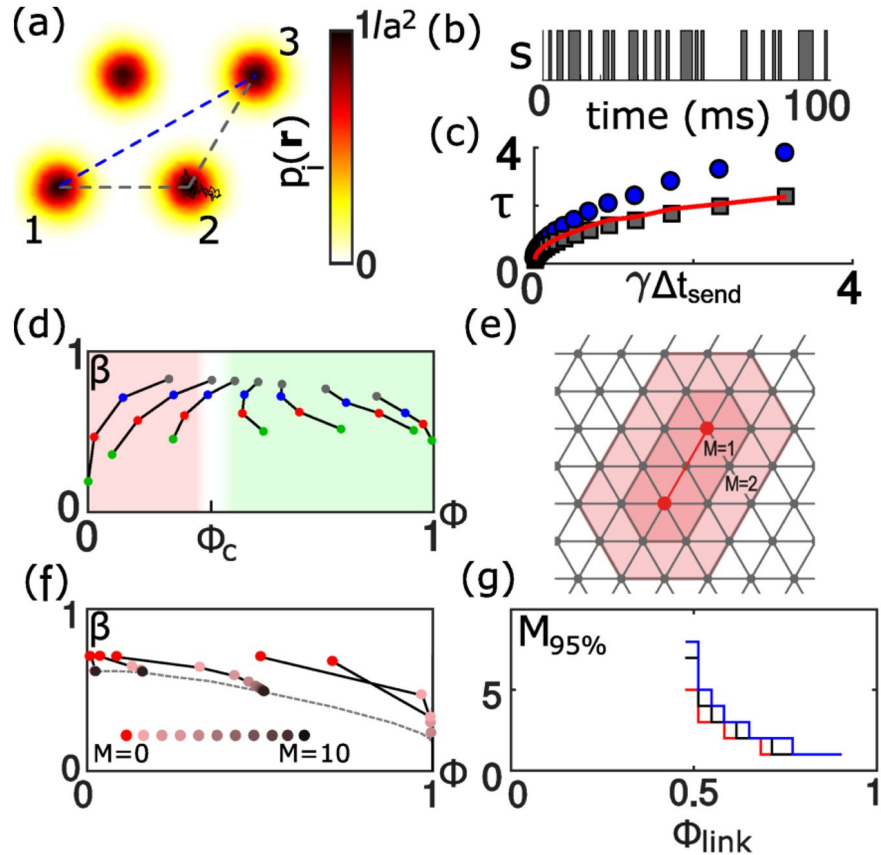

**Figure 3.** A dynamic communication network of mobile drones. (**a**) Drones diffusively explore space around reference positions on a triangular grid (color-code: probability density of positions). A communication link between two drones is active with probability $f(r)$, depending on their mutual distance $r$. (**b**) Representative time-series for the existence of an active link between two neighboring drones. (**c**) Correlation time $\tau$ for direct communication between drones 1 and 2 (gray squares), and end-to-end Bernoulliness $\beta'$ for multi-hop communication between 1 and 3 via 2 in panel (a) (blue circles). The red line shows $\beta'$ calculated from the observed $\beta$ according to RG theory for the two-link motif. (**d**) End-to-end reliability $\Phi(M_k)$ and Bernoulliness $\beta(M_k)$ for communication between source and destination nodes in network motifs $M_k$ [cf. Fig. 2(c)]. (**e**) Definition of modified flooding regions (rosé): these comprise all nodes within a distance $M$ of the shortest path (red line) between source and destination node (red dots). (**f**) Values $(\Phi_M, \beta_M)$ for multi-path routing, for increasing size $M$ of modified flooding region. (**g**) Minimal value $M_{95\%}$ to achieve an end-to-end reliability of at least 95% for three different distances $l$ between source and destination node ($l = 1$: black, 2: red, 5: blue). Parameters: $h = 20$, $\gamma/\Delta t_{send} = 100$, $D = 0.2 a^2/\gamma$, with $a$ lattice spacing, $l = 2$ in (**f**).

Our analytical theory semi-quantitatively reproduces the scaling of key network metrics with network size in agreement with full numerical simulations of large time-varying networks. To the best of our knowledge, our theory is the first to address temporal correlations of message losses in time-varying networks and the first to harness methods of renormalization group theory to address this question of both fundamental and practical interest, featuring an application of theoretical physics to computer science.

In our minimal model, we considered only two states for each link, *active* and *inactive*, which serves as illustrative example of our RG theory. The theory can be extended to the case of a finite number of discrete states, e.g. for the data transmission rate of individual links exceeding a series of thresholds, corresponding to a rated quality of service.

Future work will explore refinements of the zeroth-order renormalization theory used here, e.g. cluster methods[1–3], and generalizations to arbitrary Markov chains. We anticipate that percolation theory can guide the design of future wireless communication and control protocols that explicitly take into account temporal correlations of message losses.

### Acknowledgements

The authors are supported by the DFG through the Excellence Initiative by the German Federal Government and State Government (cluster of excellence cfaed). We thank all members of the Biological Algorithms group, in particular Justus Kromer, for stimulating discussions.

### Author Contributions

All authors jointly conceived the project, played an active role in formulating the model, and wrote the manuscript together. J.K. and B.M.F. developed the theory and performed simulations.

### Additional Information

**Supplementary information** accompanies this paper at https://doi.org/10.1038/s41598-018-25363-2.

**Competing Interests:** The authors declare no competing interests.

**Publisher's note:** Springer Nature remains neutral with regard to jurisdictional claims in published maps and institutional affiliations.